\documentstyle[preprint,aps]{revtex}
\tighten
\draft
\begin{document}
\title{
Mesoscopic Spin-Magnetism }
\author{H. Mathur,$^{(1)}$ M. G\"{o}k\c{c}eda\u{g},$^{(2)}$ and
A. Douglas Stone $^{(2)}$}
\address{$^{(1)}$AT\&T Bell Laboratories,
600 Mountain Ave., Murray Hill, New Jersey 07974-0636\\
$^{(2)}$Applied Physics, Yale University, P.O. Box 2157, New Haven,
CT 06520}

\date{}
\maketitle

\begin{abstract}
We investigate the spin-magnetism of mesoscopic metallic grains.  In
the average response of an ensemble of grains there are corrections to
macroscopic behaviour due to both spectral fluctuations and
electron-electron interactions. These corrections are a non-linear
function of the magnetic field. Their temperature dependence is
calculated numerically and analytically. An experiment is proposed to
measure the unknown interaction coupling constant in the cooper
channel. For a single sample the magnetization is found to fluctuate
reproducibly about the mean. These fluctuations directly probe the
energy level statistics.

\end{abstract}



In the early days of quantum statistical mechanics it was realized by
 Fr\"{o}hlich that the thermodynamic properties of an ideal fermi gas
in a {\it finite} box would change qualitatively from its macroscopic
behavior for temperatures ($kT$) much smaller than the mean level
spacing ($\Delta_{\epsilon}$) \cite{Frolich}.  Fr\"{o}hlich predicted
that the large level-spacing (compared to kT) would lead to exponential
temperature dependence of thermodynamic quantities $ \sim \exp( -
\Delta_{\epsilon}/kT ) $ in normal metals in contrast to the weak
T-dependence found in bulk metals.  Subsequently Kubo \cite{Kubo} and
Gorkov and Eliashberg \cite{Gorkov} took into account the {\it
fluctuations} in the level spacing of metallic grains and found that
upon averaging over an ensemble of grains the exponential dependences
were softened to power laws which depended on the precise level
statistics.  Considerable effort has been expended in experimental
searches for such finite size corrections to thermodynamics in clusters
of small metallic particles \cite{Halperin} using NMR and other
techniques.

Recent work in mesoscopic physics has pointed out that there is another
relevant energy scale for the thermodynamics of small particles, the
Thouless energy $E_c=\hbar /\tau_D$, the inverse time to diffuse across
the sample.  In a micron-size metal film $E_c \sim 1K \sim 10^4
\Delta_{\epsilon}$ so it is straightforward to perform experiments
under the conditions $E_c \gg kT \gg \Delta_{\epsilon}$, which we will
term the mesoscopic regime.  Work on orbital magnetism has shown that
in the mesoscopic regime corrections to the thermodynamics are
determined by the {\em long-range} spectral fluctuations, not the level
spacing distribution.  However the recent work on mesoscopic
thermodynamics has tended to focus on the orbital magnetization and the
associated persistent currents and has not made contact with the work
cited above which focused on spin magnetization (and specific heat).
That early work ignored the effect of long-range spectral fluctuations
and found exponential convergence to macroscopic behavior for $kT \gg
\Delta_{\epsilon}$\cite{Denton}.

In this Letter we determine the effect of spectral fluctuations on the
{\it spin magnetization}.  The spin magnetization of course contributes
to the experimentally measured magnetization and must be understood if
it is to be separated from the orbital magnetization.  Its quantum
corrections have been well-studied in the microscopic ($ kT \ll
\Delta_{\epsilon} $) and macroscopic regimes \cite{Halperin,Efetov}.
We find that in the mesoscopic regime the quantum corrections to the
average spin magnetization are not exponentially small in
$kT/\Delta_{\epsilon}$ and can be comparable to the corrections to the
orbital magnetization.  There are corrections due both to interaction
effects and to spectral fluctuations.  Experiments that probe the
thermodynamics of mesoscopic systems have been performed on both
individual samples\cite{Webb} and arrays \cite{Levy} and thus both the
mean and variance of the magnetization are in principle measurable.
Our theory indicates that the interaction corrections will give the
dominant contribution to the mean while the long-range spectral
fluctuations will be measurable in the variance, providing a direct
probe of Wigner-Dyson level statistics \cite{Mehta}.  The interaction
corrections to the mean will allow the determination of a crucial
interaction constant which also controls the size of the persistent
current.

Initially we neglect electron-electron interactions and calculate the
corrections to the Pauli susceptibility due to spectral fluctuations in
the mesoscopic regime.  In typical experiments, where the sample
dimensions are of order a micron or smaller, the thermodynamic
properties should correspond to the canonical ensemble  (CE) since the
electron number N on each specimen is fixed by charge
neutrality\cite{Kubo}.  A standard procedure in statistical mechanics
is to replace a canonical average with the average in the {\it
equivalent grand canonical ensemble} (EGCE), which is a grand canonical
ensemble (GCE) with the chemical potential adjusted to have an average
number equal to N.  Imry pointed out that this adjustment led to new
quantum interference contributions to thermodynamics proportional to
the density of states (DOS) fluctuations \cite{Imry}.  These average to
zero in macroscopic systems but become important in mesoscopic systems;
hence Imry's approach has become the standard technique for calculating
mesoscopic corrections to thermodynamics \cite{Altshuler}.  We note
however that the CE and the EGCE are {\it only} equivalent in the limit
$N \gg 1$ and $ kT \gg \Delta_{\epsilon}$ \cite{Reif}.  However in this
work we find by direct numerical simulations that the finite-size
deviations between the CE and EGCE are negligible in the mesoscopic
regime; so we will proceed using Imry's method.  In this method the
leading correction to the free energy of the EGCE is given by
\cite{Altshuler} \begin{equation} \delta F_{N} = \frac{
\Delta_{\epsilon} }{ 2 } \int_{0}^{\infty} d E \int_{0}^{\infty} d E'
< \delta \rho( E ) \delta \rho( E' ) >  f(E) f( E').  \end{equation}
Here $ f(E) $ denotes the fermi function, $\delta \rho (E)$ is the
fluctuating part of the DOS, and $ < \ldots > $ denotes an average over
impurity realizations.

In order to compute the magnetization corrections using Eq. (1) we must
obtain the magnetic field dependence of $\delta F_{N}$.  Here we focus
on systems with negligible spin-orbit interaction; in this case the DOS
fluctuation may written as \begin{equation} \delta \rho_{0} ( E +
\Delta_{z}/2 ) + \delta \rho_{0} ( E - \Delta_{z}/2 ).  \end{equation}
Here $ \Delta_{z} $ denotes the Zeeman splitting and $ \delta \rho_{0}
(E) $ denotes the DOS fluctuations for spinless electrons in a magnetic
field.  If this quantity is known the dependence of $\delta F_N$ on
$\Delta_z$ and hence the correction to the spin magnetization follows
by inserting Eq. (2) into Eq. (1) and differentiating.

It is now established by microscopic calculations that the energy
levels of disordered metallic samples are well-described by
Wigner-Dyson random matrix theory(RMT)\cite{Mehta} for energy intervals
up to $E_c$\cite{Shklovskii,Super}.  We find that the deviation from
RMT does not affect the spin magnetization when $kT \ll E_c$ and hence
we may assume that the DOS correlation function in Eq. (1) is that
given by RMT.  This correlation function decays slowly as $ 1/ \beta [
\pi(E-E') ]^{2} $, for $|E-E'|$ much larger than
$\Delta_{\epsilon}$\cite{Shklovskii,osc}.  Here $ \beta = 1,2$
depending on whether or not the orbital effects of the magnetic field
are sufficient to break time reversal symmetry.  Using this form in Eq.
(1) and differentiating one finds \begin{equation} \Delta M = - \mu_{e}
\frac{\Delta_{\epsilon}}{\pi^{2} \beta } (2 \pi kT) \sum^{ {\rm cutoff}
}_{\nu > 0 } {\rm Re} \left\{ \frac{ i \nu }{ (\nu + i \Delta_{z} )^{3}
} \right\} \end{equation} where $ \mu_{e} $ denotes the electron's
magnetic moment, $ \nu $ is a bosonic Matsubara frequency, and the
energy integrals have been converted to Matsubara sums.  This result is
insensitive to the cut-off of the sums in the regime of interest, $E_c
\gg kT \gg \Delta_{\epsilon}$.  The sum as a function of $\Delta_z$ is
easily evaluated numerically, and analytically in the limits of small
and large Zeeman splitting.  \begin{eqnarray} \Delta M  &  =  \frac{
3}{ \pi^{4} \beta } \zeta(3) \mu_{e} ( \Delta _{\epsilon} \Delta_{z} )
/ (kT)^{2}  & \; \; \; \Delta_{z} \ll kT ; \nonumber \\
	 &  = \frac{1}{2 \pi^{2} \beta } \mu_{e} ( \Delta_{\epsilon} /
\Delta_{z} ) & \; \; \; \Delta_{z} \gg kT.  \end{eqnarray} At low field
the correction to the magnetization is linear and decreases as
$(\Delta_{\epsilon}/kT)^{2} $; thus as expected it is not exponentially
small as found when long-range energy level fluctuations are completely
neglected.  Furthermore, the magnetization is {\it non-linear} at high
fields, for Zeeman splittings of the order of $kT$. The Pauli
magnetization, $M_{P} = 2 \mu_{e} \Delta_{z}/\Delta_{\epsilon} $,  by
contrast remains strictly linear at all relevant fields.  Thus despite
its small size compared to $M_P$, the correction may be distinguished
by its non-linear susceptibility \cite{Levy}.  At $ T \sim 0.1$ K the
field scale at which the non-linearity sets in is $B \sim 0.1$ T, so it
occurs at experimentally accessible parameter values.  To emphasize
that the (non-interacting) spin magnetization correction probes the
long-range level fluctuations one can easily evaluate it for the
(Poisson) case of uncorrelated levels.  Since the fluctuations are
enhanced one finds a strongly enhanced $\Delta M \sim
(\Delta_{\epsilon}/kT) M_P$.  It is possible that this behavior may be
observable in appropriate ballistic semiconductor microstructures, but
this question requires further study.

We now discuss the variance of the magnetization (which arises due to
quantum mechanical coherence).  Following Imry's method it can be shown
that chemical potential adjustment does not affect the variance to
second order in $ \delta \rho $, so one may employ the more convenient
GCE.  Using Eq. (2) the variance of the magnetization is given by
\begin{equation} \delta M^{2} = < \left( \int d E \sum_{ \alpha = \pm}
\alpha  \delta \rho_{0} ( E + \frac{\alpha \Delta_{z}}{2} ) f(E)
\right)^{2} >, \end{equation} so again the mesoscopic corrections are
determined by the correlation function of the DOS.  Using the RMT
correlation function in Eq. (5) leads to \begin{eqnarray} \delta M^{2}
& = \frac{ 6 }{ \pi^{4} \beta } \zeta(3) \mu_{e}^{2} ( \Delta_{z} /
kT)^{2}  & \; \; \; \Delta_{z} \ll kT; \nonumber \\
 & = \frac{ 1 }{ \pi^{2} \beta } \mu_{e}^{2} \ln ( \Delta_{z} /kT) & \;
\; \; \Delta_{z} \gg kT.  \end{eqnarray} Note that the
root-mean-squared magnetization fluctuation is parametrically larger
than the correction to the mean.  It grows linearly with Zeeman
splitting at low field and slows to a logarithmic increase at high
fields.  The behavior for $kT \gg \Delta_z$ follows from our previous
result for $\Delta M$ within Imry's method; if one expands Eq. (5) to
2nd order in $\Delta_z$ and integrates by parts one finds $ \Delta M
/M_P = \delta M^2/ M_P^2$, independent of the nature of the level
statistics.  It thus follows that in general $\delta M/ \Delta M =
(M_p/\Delta M)^{1/2} \gg 1$, and the magnetization fluctuations at low
field will always be much larger than the correction to mean. The
saturation of $\delta M^2$ to a logarithmic increase at high-fields
($\Delta_z \gg kT$) also has a simple interpretation.  At T=0 the GCE
magnetization literally counts the number of levels (at zero splitting)
in an energy window of width $ \Delta_{z} $ centred at $\epsilon_f$.
Thus $\delta M^2$ measures the variance of that number.  It is a famous
result of RMT that this variance increases logarithmically with the
size of the interval \cite{Mehta} (as opposed to the linear increase
for uncorrelated levels); hence the logarithmic dependence on
$\Delta_z$.

By the familiar ``ergodic hypothesis'' of mesoscopic physics
\cite{Doug} we expect the statistical fluctuations of $M$ to manifest
themselves as fluctuations of the magnetization of a given specimen as
a function of external parameters (e.g. fermi energy or magnetic
flux).  However simply varying magnetic field is not a good method as
this changes $\Delta_z$ as well.  With a 2D mesoscopic sample one can
envision varying the tilt angle of the field with respect to the plane
of the sample, hence varying the magnetic flux at fixed $\Delta_z$.
The main difficulty with such an experiment is the small absolute size
of the effect, which is of order a Bohr magneton per sample.  However
NMR has the sensitivity to measure such small effects (roughly 1-10
{\it ppm} for a micron-size film at $B \sim 1T$) if appropriate samples
can be obtained.  Here the mesoscopic size scale offers entirely new
possibilities as one can imagine fabricating arrays of metal samples
lithographically (as was done to measure persistent currents
\cite{Levy}) and achieving much greater homogeneity than in
conventional NMR of small particles.  The mesoscopic fluctuations would
appear as a broadening of the NMR line which increases with decreasing
temperature \cite{Efetov}.

As noted above, Imry's method assumes that the EGCE, which uses fermi
functions with an adjusted chemical potential yields, the same
thermodynamic properties as the true canonical ensemble (which can
never be exactly represented by a fermi function), at least in the
mesoscopic regime ($E_c \gg kT \gg \Delta_{\epsilon}$).  Since it has
been shown that the CE and EGCE have substantially different
thermodynamic properties when $kT < \Delta_{\epsilon}$\cite{Halperin},
it seemed necessary to test their equivalence in the mesoscopic
regime.  Moreover Imry's formula (Eq. (1)) is the first term in an
expansion involving successively higher level correlation functions;
therefore its quantitative accuracy for a particular range of values of
$kT/\Delta_{\epsilon}$ needs to be tested.

To do this we have calculated the exact ensemble-averaged CE and EGCE
spin magnetization numerically.  The ensemble of level-sequences was
generated by direct diagonalization of $10^3$ $1000 \times 1000$
gaussian orthogonal ($\beta=1$) random matrices and appropriate
``unfolding'' of the levels obtained\cite{Mehta,Bohigas}.  Evaluation
of the CE partition function (for a given level sequence) by brute
force summation is not possible for $kT \gg \Delta_{\epsilon}$ due to
the enormous number of relevant states. To bypass these difficulties we
use the Darwin-Fowler representation of the CE partition
function\cite{Darwin}.  In this approach an infinite (Darwin-Fowler)
polynomial is defined whose coefficients are precisely the partition
function for different values of $N$.  Each coefficient can be
extracted (in principle) by an appropriate contour integral.
Evaluation of this integral by steepest descent yields the EGCE
partition function \cite{Darwin}.  We have shown that it is possible to
represent the magnetization directly as a Darwin-Fowler type contour
integral, which can be evaluated numerically for sequences of 500
levels without difficulty.  Exact numerical calculations within the
EGCE are straightforward as one may use the fermi-function and simply
adjust the chemical potential appropriately for each level-sequence.
This procedure includes the higher order correlations neglected in Eq.
(1).

The results of these calculations for the magnetization are shown in
Fig. 1.  They may be summarized as follows: 1) When $kT >
\Delta_{\epsilon}$ the CE and EGCE averages agree almost perfectly.
Hence the EGCE is an excellent approximation in the mesoscopic regime
\cite{Kamenev}.  2) The agreement between the numerical and analytic
calculations of $\delta M^2$ is very good whether the variance is
calculated in the (unadjusted) GCE or the EGCE.  3) The analytic
formula for the mean correction $\Delta M$ and the numerical
calculations differ substantially in the interval $\Delta_{\epsilon}<
kT < 10 \Delta_{\epsilon}$ (roughly by a factor three).  It is also not
possible to fit the numerical results to the quadratic T-dependence
predicted (although this may be due to the breakdown of the accuracy of
our numerics at the highest values of $kT$).  This shows that there are
substantial corrections to Eq. (1) in the range of parameters where the
mesoscopic corrections are largest ( $ \Delta_{\epsilon} < kT \leq 10
\Delta_{\epsilon} $).

Finally we include the effects of interaction on the mean spin
magnetization using perturbation theory.  It is known that these
effects are important for persistent currents in disordered metals.
The relevant diagrams for the spin magnetization are already known in
the literature \cite{Fukuyama}, only their behavior for $kT < E_c$ had
not been explored.  We find \begin{equation} \Delta M_{int} = ( F +
\lambda_{c} ) kT \sum_{\Omega} \sum_{\nu} {\rm Re} \left\{ \frac{ - i
\nu }{ (\nu + i \Delta_{z} + E_{c} \Omega^{2} )^{2}
 } \right\} \end{equation} where $F,\lambda_c$ are coupling constants
in the diffuson, cooperon channels which we will regard as parameters
to be determined experimentally.  Here $ \Omega^{2} $ denotes the
eigenvalues of $ L^{2} \nabla^{2} $ ($L$ is a typical sample dimension)
with Neumann boundary conditions.  $\Omega^2 = 0$ is always an
eigenvalue and it is easy to show that this term dominates the sum when
$kT < E_c$. Keeping only this term and performing the Matsubara sums
yields \begin{eqnarray} \Delta M_{int} & = ( F + \lambda_{c} ) \frac{
1}{\pi^{2} } \zeta(2) (\Delta_{z}/kT)
 & \; \; \; \Delta_{z} \ll kT ; \nonumber \\ = & F + \lambda_{c} & \;
 \; \; \Delta_{z} \gg kT.  \end{eqnarray} This correction is
parametrically larger than that predicted by the non-interacting theory
(Eq. 4) if the dimensionless interaction constants are not too small.
Note that $\Delta M_{int}$ is linear at low field ($ \Delta_{z} \ll kT
$) and saturates at high field ($ \Delta_{z} \gg kT $). It is therefore
highly non-linear for fields around $ \Delta_{z} \sim kT $. In this
respect its response mimics that of an isolated spin which may
complicate its experimental determination.  However because it involves
phase-coherent diffusion the interaction correction is very
anisotropic.  Eq. (8) assumes that the cooperon term is not suppressed;
the magnetic flux at which this term is suppressed is the flux quantum
$h/e$ which corresponds to a normal magnetic field of order $10^2$ G.
As noted above, if samples are fabricated as small metal films the flux
may be varied by tilting the sample allowing a measurement of the term
proportional to $\lambda_c$.  This would be an important result of such
an experiment as $\lambda_c$ controls the size of the persistent
current and its value is only measurable very indirectly in
non-superconducting materials \cite{Ambegaokar}.  If such a
measurement, e.g. in copper, agreed with the measured amplitude of the
persistent current \cite{Levy} it would provide strong evidence that
the interaction correction is the relevant one in that case as well.
Finally, in the case of persistent currents the interacting and
non-interacting theories both give exponential temperature-dependence
whereas for the spin magnetization different power-laws are predicted
(see Eqs. (4),(8)) which may make it easier to experimentally
distinguish the two effects.

To summarize, we have studied the field and temperature dependences of
mesoscopic corrections to macroscopic spin-magnetism for grains of
normal metal. Measurement of the correction for an ensemble of grains
would provide an independent estimate of an important interaction
coupling constant which is difficult to measure in non-superconducting
metals.  Single samples are found to exhibit magnetization fluctuations
that depend only on the external parameters, the applied field and
temperature, not on sample properties such as size, impurity density or
fermi energy.

We thank H. Baranger and S. Barrett for helpful comments.  This work
was partially supported by NSF grant DMR-9215065.

\figure{ Comparison of the temperature dependence of the average
magnetization at low field, calculated in the different ensembles:
Squares represent the numerical CE result; triangles, the numerical
EGCE result; and the smooth curve, the analytical result of Eq (4).
Inset: The same results on a log-log plot. The magnetization
fluctuations (normalized by $ M_{P}^{2} $) are also shown.  With this
normalization, the analytical curve for the fluctuations coincides with
that for the average (see text); circles represent the numerical GCE
result. \label{Fig. 1}}



\begin{references}
\bibitem{osc} This form is an approximation to the exact
two-level correlation functions of the Wigner-Dyson ensembles which can
be shown to apply when $kT \gg \Delta_{\epsilon}$.  Numerical
evaluation of Eq. (1) employing the exact two-level correlation
functions (which exhibit a rapid oscillation on the scale of
$\Delta_{\epsilon}$) yields a negligible correction to Eq. (4) down to
$kT \approx \Delta_{\epsilon}$.
\end{references}
\end{document}